\documentclass[12pt]{article}

\begin{document}

\begin{center}

\LARGE{\bf Determination of a magnetization parameter of the parsec-scale AGN jets}

\vspace*{.5cm}

\large{\it V. S. Beskin$^{}$\thanks{E-mail:
beskin@lpi.ru}, Y. Y. Kovalev$^{1}$ and E. E. Nokhrina$^{2}$}

\vspace*{.5cm}

\small{$^{1}$P.N.Lebedev Physical Institute, Leninsky prosp., 53, Moscow, 119991, Russia\\
$^{2}$Moscow Institute of Physics and Technology, Dolgoprudny,
141700, Russia}

\end{center}





\begin{abstract}

The observed shift of the core of the relativistic AGN jets as a function of frequency allows 
us to evaluate the number density of outflowing plasma and, hence, the multiplicity parameter 
$\lambda = n/n_{\rm GJ}$. The value $\lambda \sim 10^{13}$ obtained from the analysis of more 
than 20 sources shows that for most of jets the magnetization parameter $\sigma \sim 10$--$100$. 
Since the magnetization parameter is the maximum possible value of the Lorentz factor of the 
relativistic bulk flow, this estimate is consistent with the observed superluminal motion.

\end{abstract}


\section{Introduction}

One of the most important parameters in magneto-hydrodynamic (MHD) models of relativistic 
jets is the dimensionless multiplicity parameter $\lambda = n/n_{\rm GJ}$, which is defined 
as the ratio of the number density $n$ to the so-called Goldrech-Julian (GJ) number density 
$n_{\rm GJ} = \Omega B/2 \pi c e$ (i.e., the minimum concentration required for the screening 
of the longitudinal electric field in the magnetosphere). It is important that the multiplicity 
parameter associates with the magnetization parameter $\sigma$, which determines the maximum
possible bulk Lorentz-factor of the flow, which can be achieved (Beskin, 2010)
\begin{equation}
\sigma \approx \frac{1}{\lambda}\left(\frac{W_{\rm tot}}{W_{\rm A}}\right)^{1/2}.
\label{newsigma}
\end{equation}
Here $W_{\rm A} = m_{\rm e}^{2}c^{5}/e^{2} \approx 10^{17}$ erg/s, and $W_{\rm tot}$ is the 
total energy losses of the compact object. If the inner parts of the accretion disc are hot 
enough, these electron-positron pairs can be produced by two-photon collisions, the photons 
with sufficient energy delivering from the inner parts of the accretion disk (Blandford \& 
Znajek, 1977). In this case \mbox{$\lambda \sim 10^{10}$--$10^{13}$,} and the magnetization 
parameter $\sigma \sim 10^{2}$--$10^{3}$. The second model takes into account the appearance 
of the region where the GJ plasma density is equal to zero due to the GR effects that corresponds 
to the outer gap in the pulsar magnetosphere (Beskin et al, 1992, Hirotani \& Okamoto, 1998). 
This model gives $\lambda \sim 10^{2}$--$10^{3}$,  and   $\sigma \sim 10^{10}$--$10^{12}$.

\section{The method}

To determine the multiplicity parameter $\lambda$ and the  magnetization parameter $\sigma$ 
one can use the dependence on the visible position of the core of the jet from the observation 
frequency (Gould, 1979, Howatta et al, 1979, Marscher, 1983, Lobanov, 1998, Hirotani, 2005, 
Gabuzda et al, 2008). This effect is associated with the absorption of the synchrotron photon 
gas by relativistic electrons in a jet. The apparent position of the nucleus is determined by 
the distance at which for a given frequency the optical depth reaches unity. Such measurements 
were performed by Sokolovsky et al, (2011) for 20 objects. Observations at nine 
frequencies allowed to approximate the apparent position of the nucleus as a function of frequency
\begin{equation}
r_0 - r = \xi -\eta \left(\frac{\nu}{\rm GHz}\right)^{-1},
\label{deltar}
\end{equation}
where $r_{0}$ is the position of the bright area of the emission, $r$ is the apparent position 
of the nucleus in mas, and $\nu$ is the frequency. Here, the quantities $\xi$ (in mas) and $\eta$ 
(in \mbox{mas $\cdot$ GHz)} are the measured parameters of this approximation. Knowing this 
dependence and assuming the equipartition of energy between the particles and the magnetic field, 
one can write down
\begin{eqnarray}
\lambda = 2.6 \times 10^{12} \,
\left(\frac{\eta}{\rm mas \, GHz}\right)^{3/2}
\left(\frac{W_{\rm tot}}{10^{47} {\rm erg/s}}\right)^{-1/2} \nonumber \\
\left(\frac{D_{\rm L}}{1 \, {\rm GPc}}\right)^{3/2}
\sqrt{\frac{K}{\gamma_{\rm in}^2}}
\frac{1}{\sqrt{\chi} \sin\varphi \, \delta \, (1+z)^2}.
\label{newsigma1}
\end{eqnarray}
Here $D_{\rm L}$ (Gpc) is the object distance, $\chi$ (rad) is the opening angle of ejection, 
$\varphi$ (rad) is the angle of view, $\delta$ is the Doppler factor, $z$ is the red-shift, 
and $K$  is the dimensionless function of the minimum and maximum Lorentz-factor of electrons 
in their power-law distribution in energy (Marscher, 1983). Thus, for the 20 objects for which 
parameter $\eta$ was measured, we can estimate the magnetization parameter $\sigma$.


\begin{table}
\caption{The apparent frequency-dependent shift of the nuclei, the multiplicity 
parameter $\lambda$ and the magnetization parameter $\sigma$.}  

\vspace*{.5cm}

\centering
\begin{tabular}{|c|c|c|c|c|}
  \hline
object   & $\eta$(mas GHz)  &     $z$   & $\lambda$($10^{13}$) & $\sigma$ \\
\hline
0148+274    & 3.4    & 1.3    & 21.0    & 4.8 \\
\hline
0342+147    & 1.0    & 1.6    &  3.7    & 27 \\
\hline
0425+048    & 2.2    & 0.6    &  6.5    & 15 \\
\hline
0507+179    & 1.7    & 0.4    &  3.6    & 28 \\
\hline
0610+260    & 3.6    & 0.6    & 14.5    & 6.9 \\
\hline
0839+187    & 2.3    & 1.2    & 11.2    & 9.0 \\
\hline
0952+179    & 1.4    & 1.5    &  5.9    & 16 \\
\hline
1004+141    & 2.4    & 2.7    & 14.3    & 7 \\
\hline
1011+250    & 2.1    & 1      &  9.0    & 11 \\
\hline
1049+215    & 1.8    & 1.3    &  7.8    & 12 \\
\hline
1219+285    & 2.5    & 0.1    &  6.2    & 16 \\
\hline
1406-076    & 1.2    & 1      &  3.9    & 26 \\ 
\hline
1458+718    & 2.4    & 1      & 11.3    & 8.9 \\
\hline
1642+690    & 1.9    & 0.8    &  6.5    & 15 \\
\hline
1655+077    & 1.5    & 1      &  5.4    & 19 \\ 
\hline
1803+784    & 1.1    & 0.7    &  6.6    & 15 \\ 
\hline
1830+285    & 2.8    & 0.6    &  9.8    & 10 \\
\hline
1845+797    & 2.3    & 0.1    &  0.5    & 199  \\
\hline
2201+315    & 3.3    & 0.3    &  6.5    & 15 \\
\hline
2320+506    & 1.3    & 1      &  3.8    & 27 \\
\hline
\end{tabular}
\label{table1} 
\end{table}

In Table 1 we present the obtained results. 
Here $\eta$ are taken from observations of 20 objects Sokolovdsky et al, (2011), the red-shifts 
$z$ are taken from Kovalev et al, (2008), and the distance to the object was determined from 
the redshift. For the five objects for which the red-shift is unknown, we took $z$ = 1. As 
the half-opening angle, the angle between the jets and the line of sight (viewing angle) and 
Doppler factors were taken typical values: $\delta = 6^{\circ}$, $\chi = 9^{\circ}$, 
$\varphi = 2^{\circ}$, except for objects 1803+784 and 2201+315. Doppler factor and the angle 
of view for the source  1803+784 was taken from Homatta et al, (2008), and the half opening 
angle of jet of this object was taken from  Jorstad et al, (2005). Doppler factor and viewing 
angle for 2201+315 is taken from Jorstad et al, (2005). In addition, we have put for the full 
power losses $W_{\rm tot} = 10^{47}$ erg/s, which corresponds to the Eddington luminosity for 
the central object mass $10^{9} \, M_{\odot}$.

\section{Conclusion}

The obtained values of the multiplicity parameter $\lambda$ of the order $10^{13}$--$10^{14}$ 
are consistent with the Blandford-Znajek model. At the same time, this value corresponds to 
the concentration of particles which were found by Lobanov (1980). The magnetization parameter 
$\sigma$ of the order of 10 or several dozen is in agreement with the Lorentz factor values 
estimated by Cohen et al, (2007), from VLBI jet kinematics measurements. Additionally, for 
1803+784 the Lorentz-factor is suggested to be equal to  9.5 (Sokolovsky et al, 2011), whereas 
we found $\sigma = 10.2$. For \mbox{2201 +315} we have $\gamma= 8.1$ and $\sigma = 15.4$. In both 
cases $\gamma < \sigma$. For different types of objects (quasars, blazars, and radio galaxies) 
found by Howatta et al, (2009). the average Lorentz factors range from 2 to 17, that is about ten, 
which support our point of view as well.
Thus:
\begin{enumerate}
\item 
By measuring the apparent shift of the core jet emission as a function of frequency for 20 
objects we obtained the estimates of the multiplicity $\lambda \sim 10^{13}$, which corresponds 
to the Blandford-Znajek effective production of secondary particles (see Moscibrodzka et al, 
2011 as well).
\item
For most objects the magnetized parameter $\sigma \sim 10$, which is in good agreement with the 
observed superluminal motion.
\end{enumerate}

\section{Acknowledgments}

We would like to acknowledge D.~Gabuzda and M.~Sikora for useful discussions.
VSB thanks Uniwersytet Jagiello\'nski for hospitality. 
The work was supported by RFBR grant 11-02-02021 and the Ministry of Science and Education 
(contract No. 02.740.11.0250).

\end{document}